\documentstyle[12pt]{article}
\begin{document}
\begin{center}
{{\bf \large Sum rules for isospin centroids in pick-up reactions on
general multishell target states}}\\[12pt] 
{R. K. Bansal,  H. Sharda }\\
{\it Department of Physics, Panjab University, Chandigarh-160 014,
India.}\\[6pt] 
{and}\\
{Ashwani Kumar }\\
{\it Department of Applied Sciences, Punjab Engineering College,
Chandigarh-160 012, India.} 
\end{center}
\*
\vskip 3truecm
\begin{abstract}
\label{abs}
\baselineskip 24pt
Sum Rules equations for pick-up reactions are presented for the first time
for the energy centroids of states both for the isospin $T_< {(\equiv
T_{0}-{1\over 2})}$ and $T_>{(\equiv T_0 + {1 \over 2})}$ of the final
nucleus when a nucleon is picked up from a general multishell target state
with isospin $T_{0}$. These equations contain two-body correlation terms,
$<H^{01}>$, which, at the present moment, are difficult to handle
analytically.  These terms are managed by combining these equations with
the known stripping reactions equations. Sample applications of these
equations to experimental data are presented.
\end{abstract}
\vfill \eject
\newpage
\baselineskip 24pt
\subsection*{1. Introduction}
\label{int}
Sum rule methods have long been in use for analysing experimental data
obtained from direct transfer reactions.  The linear energy-weighted sum
rules [1,2] relate the strengths and energies of residual nucleus states
to multipole moments of the target state and the effective nucleon-nucleon
interaction.  In particular, the monopole energy weighted sum rule may
be used to obtain an expression for the energy centroid of states of the
residual nucleus.  For inequivalent transfer involving a general
multishell target state, the expression for centroid energy turns out to
be fairly simple [3]. 
\par For equivalent transfer, the expressions for isospin centroids of
states obtained via single particle addition to a target state, having
only neutrons in the transfer orbit [4],though a bit more involved, are
still amenable to useful applications [5,6]. 
\par
However, when no restriction is imposed on the occupancy of orbits in the
target state, the equations for isospin centroids, obtained by equivalent
single particle stripping [7], contain an isovector two-body correlation
term which defies simple analytical evaluation unlike the other terms
appearing in these equations. This difficulty could be overcome, if the
same stripping experiment provides complete information about the states
of the residual nucleus having $T_<$ and $T_>$ values of the isospin,
because in that case the `problem term' could be eliminated. However,
usually it is not feasible to extract this complete information from a
single stripping reaction and this limits the scope of application of sum
rules to generalized stripping situations.
\par
The aim of this article is to present, for the first time, explicit
algebraic expressions for isospin centroids of states of residual nucleus
obtained via single nucleon pick-up from a general multishell target
state. As expected, these expressions also involve the same isovector
two-body correlation term, $\langle H^{01}\rangle$, as found in the case
of stripping situation for the same target state.  It is, therefore,
possible to eliminate this term by combining the equations for isospin
centroids of the states of two residual nuclei obtained from the same
target state, one via stripping and the other via pick-up of one nucleon.
These combinations will prove much more useful as these correspond to the
actual experimental situation, namely, that states belonging to one
particular isospin $(T_<$ or $T_>)$ are usually populated by single
particle transfer reactions.  Sample applications are also presented to
illustrate this approach. 
\subsection*{2. Sum Rules and Energy Centroids}
\label{srec}
The general linear energy weighted sum rule for single particle pick-up
may be written as [2,7]
$$
\sum_{\Gamma x} (-1)^{\Gamma_0+\rho_i-\Gamma}
(2\Gamma_0+1)(2\Gamma+1)^{-1/2}
U(\Gamma_0\rho_i\Gamma_0\rho_i;\Gamma\Lambda)S^{-}_{\Gamma x}
E^{-}_{\Gamma x} 
$$
$$
=(-1)^{\Lambda}\langle n\Gamma_0 x_0 \parallel(A^{\rho_i} \times H \times
B^{\rho_i})^{\Lambda}  \parallel n \Gamma_0 x_0 \rangle \eqno(1)
$$
where we use the familiar notion of the product space with each Greek
letter representing two quantities ---one in the angular momentum space
and the other in the isospin space. Thus $\Gamma_{0}$ stands for $J_{0}$
and $T_{0}$; $\Gamma$ for J~ and~T;~ $\rho$ for j~and~$1\over 2$;
$\Lambda$ for $\kappa$~and~$\tau$; and the factors
$(2\Gamma_{0}+1)\equiv(2J_{0}+1)(2T_{0}+1)$ etc. Here $J_{0}$, j, J refer
to the angular momentum of the target state, a nucleon in orbit $\rho$ and
a residual nucleus state respectively. $T_0$,$1\over2$, T are the
corresponding isospins. The target state is $\mid n \Gamma_o x_o\rangle$
with $n$ being the number of nucleons and $x_o$ standing for all non
angular momentum quantum numbers required to specify the target state
uniquely. $A^\rho$ and $B^\rho$ are properly symmetrized and normalized
creation and destruction operators respectively for a nucleon in orbit
$\rho$. The subscript $i$ is used here for the transfer orbit.
$E^{-}_{\Gamma x}$ and $S^{-}_{\Gamma x}$ are the energy and spectroscopic
factor, respectively, of the state $\mid n-1,\Gamma x\rangle$ of the
residual nucleus with superscript ``-'' reminding us that this state is
obtained by pick-up (removal) of a nucleon from the target state.
$\Lambda \equiv \kappa \tau$ represents the rank of the operator
$(A^{\rho_i}\times H\times B^{\rho_i})^{\Lambda}$ in both angular momentum
space $(\kappa)$ and isospin space $(\tau)$. ${\kappa}$ can take the
values 0,1,2,\ldots, $2j_<~~~(j_<$ being the smaller of $j_i$ and $J_0)$
while $\tau=0$ or 1. The Hamiltonian H can be written as
$$ H=H(c.s.)+\sum_r\epsilon_{\rho_r}n_{op}(\rho_r)+H^{(2)} \eqno(2)
$$
where$ H(c.s.)$ is the contribution from closed shells (inert core) and
the second term is the effective one-body part of the Hamiltonian . The
two-body part is given by
$$
H^{(2)}=-\sum_{rstu,\gamma ;~ r\leq s,~t\leq u} {\frac{(2\gamma
+1)^{1/2}}{(1+\delta_{rs})^{1/2}(1+ \delta_{tu})^{1/2}}} W^{\gamma}_{rstu}
\{(A^{\rho_r} \times A^{\rho_s} )^{\gamma} \times (B^{\rho_t} \times
B^{\rho_u} )^{\gamma} \}^0\eqno(3) 
$$
In this expression $W^{\gamma}_{rstu}$ represents the antisymmetrized
matrix element,
\par
$\langle\left(\rho_{r}\rho_{s}\right)^{\gamma}\mid H^{(2)} \mid
\left(\rho_{t}\rho_{u}\right)^{\gamma}\rangle$, of the effective two-body
nucleon nucleon interaction. The summation indices, r,s,t,u run over all
the active orbits and are supposed to be ordered in some definite fashion.
\par
Substituting equations (2) and (3) in equation (1), performing standard
algebraic simplifications, we obtain, for the nucleon pick-up case from
the $\rho_i$ --orbit of the target state, the following expression for the
isospin centroid of the residual nucleus states.
$$
E^{-}_{T}-E^{-}(riz) =
\frac{\sum_{Jx}\left(\frac{2T_0+1}{2T+1}\right)S^{-}_{JTx}E^{-}_{JTx}}
{\sum_{Jx}\left(\frac{2T_0+1}{2T+1}\right)S^{-}_{JTx}}-E^{-}(riz) 
$$
$$
=\frac{\sum_k \left\{
<H^{00}_{ik}>_{Tar}-{\frac{f(T)}{T_0}}<H^{01}_{ik}>_{Tar} \right\}} 
{ \left\{{n_i\over 2}- {\frac{f(T)T_{0i}}{T_0}} \right\}} \eqno(4)
$$
It may be mentioned here that while evaluating various matrix elements
during the algebraic process, we have assumed a pure multishell
configuration for the target state, where, in the spirit of the low energy
approximation, the isospin,
\par 
$T_0 =T_{0z}=(N-Z)/2$. 
\par 
In the equation (4), $E^{-}{(riz)}$ is the energy of the ``Residual
Interaction Zero'' state of the final nucleus, that is, the state obtained
by assuming that the picked-up particle had no interaction with the active
nucleons in the target state. Thus
$E^{-}{(riz)}=E_{0}-{\epsilon_{\rho_i}}$, where $E_0$ is the target state
energy and $\epsilon_{\rho_i}$ is the single-particle energy of a particle
in ${\rho_i}$ orbit with respect to the inert core. The factor $f(T)$ in
the equation is given by
$$
f(T) = T(T+1)-{3 \over 4}-T_0(T_0+1)=\left\{ \begin{array}{cc}
{T_0} & {for~T=T_{>}(\equiv T_0+{1 \over 2})}\\ 
{-(T_0+1)} & {for~T=T_{<}(\equiv T_0-{1 \over 2})} 
\end{array} \right.\eqno(5) 
$$
The summation index $k$ on the right hand side of equation (4) runs over
all the active orbits in the target state and $<H^{\Lambda}_{ik}>_{Tar}$
is the expectation value, in the target state, of the two-body correlation
operator given by 
$$
H^{\Lambda}_{ik} = {1\over2}{\sum_{\gamma}(2{\gamma}+1)^{1 \over 2}
W^{\gamma}_{ikik} [ \{ (A^{\rho_k} \times A^{\rho_i})^{\gamma} \times
B^{\rho_k} \}^{\rho_i} \times B^{\rho_i} ] ^{\Lambda} } \eqno(6) 
$$
\noindent For the isoscalar correlation operator, the expectation value
turns out to be 
$$
<H^{00}_{ik}>_{Tar} = -{1 \over 2} (1+\delta_{ik})E^{(2)}_{Tar}(i-k)
\eqno(7) 
$$
where $E^{(2)}_{Tar}(i-k)$ is the contribution to the target state energy
from two-body interaction of active nucleons in the $i$th orbit with those
in the $k$th orbit. 
\par
$<H^{01}_{ik}>_{Tar}$ in equation (4) is the expectation value of the
isovector two-body correlation operator, in the target state, which cannot
be evaluated analytically or simply in the same manner as
$<H^{00}_{ik}>_{Tar}$. 
\par
The denominator on the right hand side in equation (4) is given by the
non-energy weighted sum rules [8] in terms of the occupancy $n_{i}$ of the
transfer orbit $\rho_i$ in the target state, and the partial isospin
contribution, $T_{0i}$, of the $\rho_i$ orbit ; using these results, we
can write for the $T_>(\equiv T_{0}+{1\over2})$ and $T_<( \equiv T_{0}-{1
\over 2})$ centroids 
$$
E^{-}_{T_>}-E^{-}(riz) = \frac{\sum_{k} \{ <H^{00}_{ik}>_{Tar}-
<H^{01}_{ik}>_{Tar} \} } {<{protons}>_{\rho_i}} \eqno (8)
$$
and\\
\noindent
$E^{-}_{T_<}-E^{-}(riz)
$
$$
=\frac{\sum_k{ \{ <H^{00}_{ik}>_{Tar} +
\left(\frac{T_0+1}{T_0}\right)<H^{01}_{ik}>_{Tar}\} }}
{<{neutrons}>_{\rho_i} + {\frac{1}{2T_0}}
\{<{neutrons}>_{\rho_i}-<{protons}>_{\rho_i}\}}\eqno(9) 
$$
The term $<H^{01}_{ik}>_{Tar}$ occuring in these equations defies simple
analytical evaluation, thus limiting the scope of application of any of
these individually. If data are available for pick-up reactions, on the
same target leading to both $T_<$ and $T_>$ states, then the equations (8)
\& (9) may be suitably combined to eliminate this term. However, in the
experimental situation, usually, such is not the case. More often, we find
one stripping and one pick-up reaction on the same target, in each case
providing reliable information about states having a particular T-value.
Therfore, it is often convenient to eliminate the term $<H^{01}>$, of
course, depending on the particular situation, by combining one of the
pick up equations (8) \& (9) with one of the equations, for T-centroids
from single particle stripping reactions[7]. For convenience we reproduce
these below :\\
\noindent
$
E^{+}_{T_>}-E^{+}(riz)
$
$$
=\frac{\sum_{k}{\{p^{+}_{T_>}(i-k)+(N_i-
\delta_{ik})q^{+}_{T_>}(k)\overline{W}^{T=1}_{ik} + (N_{i}+
\delta_{ik})r^{+}_{T_>}(k)\overline{W}^{T=0}_{ik} \}}}
{<{neutron (holes)}>_{\rho_i}}\eqno(10)
$$
and\\
\noindent
$
E^{+}_{T_<}-E^{+}(riz)
$
$$
= \frac{\sum_k{ \{p^{+}_{T_<}(i-k) + (N_i-
\delta_{ik})q^{+}_{T_<}(k)\overline{W}^{T=1}_{ik}
+(N_i+ \delta_{ik})r^{+}_{T_<}(k) \overline{W}^{T=0}_{ik} \}}}{<{proton
(holes)}>_{\rho_i}+{\frac{1}{2T_0}}\{<{proton
(holes)}>_{\rho_i}-<{neutron (holes)}>_{\rho_i}\}}\eqno(11)
$$
\noindent In these equations
$$
p^{+}_{T}(i-k)~=~<H^{00}_{ik}>_{Tar}+\frac{f(T)} {T_{0}}
<H^{01}_{ik}>_{Tar} \eqno(12)
$$
$$
q^{+}_{T}(k)~=~\frac{3n_{k}}{4}~+~\frac{f(T)T_{0k}}{2T_{0}} \eqno(13)
$$
$$
r^{+}_{T}(k) ={n_{k}\over 4} -{\frac{f(T)T_{0k} }{2T_0}}\eqno(14)
$$
$$
N_{i} = 2j_i +1 \eqno(15) $$ and $$ E^{+}(riz)=E_0 +{ \epsilon_{ \rho_i}
}\eqno(16) 
$$
$\overline{W}^{T=1}_{ik}$ and $\overline{W}^{T=0}_{ik}$ appearing in
equations (10) \& (11) are the average two-body interaction energies in
the isotriplet and isosinglet states, respectively, of one nucleon in the
$i$th orbit and the other in the $k$th orbit.  These are given by
$$
\overline{W}^{T}_{ik} =\frac{ \sum_{J} (2J+1) W^{JT}_{ikik} }{\sum_{J}
(2J+1)} \eqno(17) 
$$
The symbol $n_{k}$ in equations (13) \& (14) is the number of nucleons in
the $k$th active orbit ofthe target state whereas $T_{0k}$ is, in some
sense, the partial contribution of these nucleons towards the isospin 
of the target state.
\subsection*{3. Applications and Discussion }
\label{ad}
As discussed in the previous section, a combination of any of the
equations (8) \& (9) with one of (10) and (11) helps us to eliminate the
term $<H^{01}_{ik}>_{Tar}$. Assuming a pure multishell configuration for
the target state with $T_0 =T_{0z}={(N-Z)/2}$, we can easily evaluate the
factors $q^{+}_{T}$ and $r^{+}_{T}$.  The values of $E^{-}(riz)$,
$E^{+}(riz)$ and $<H^{00}_{ik}>_{Tar}$ can be computed using the Binding
Energy Tables [9]. Obtaining the values of $E^{-}_{T}$ and $E^{+}_{T}$
from experimentally measured energies and spectroscopic factors in one
pick-up and one stripping reaction experiments on the same target state,
we can extract the values of average effective interaction parameters
$\overline{W}^{T=1}$ and $\overline{W}^{T=0}$ by making least-squares
fits. 
\par In the present study we have limited ourselves to reactions involving
the transfer (both pick-up and stripping) of a nucleon to the
$1f_{7\over2}$ orbit with targets having $1f_{7\over2}$ as the only active
shell outside inert core $^{40}$Ca.  The reaction data used in setting up
the equations for $\overline{W}^{T=1}_{f_{7\over2}f_{7\over2}}$ and
$\overline{W}^{T=0}_{f_{7\over2}f_{7\over2}}$ are listed in Table 1.  The
seventeen linear equations, so formed, have been used to obtain the best
fitted values.  The values of
$\overline{W}^{T=1}_{f_{7\over2}f_{7\over2}}$ and
$\overline{W}^{T=0}_{f_{7\over2}f_{7\over2}}$ so 
obtained are compared with the previous results in Table 2.
\par 
In our previous calculations of average effective interaction parameters,
we had to restrict ourselves to targets having only active neutrons
because in that case, the isovector two-body correlation term
$<H^{01}_{ik}>_{Tar}$ happens to be equal to isoscalar correlation
$<H^{00}_{ik}>_{Tar}$ due to certain specific isospin constraints. The
equations for isospin centroids obtained via single nucleon pick-up,
presented in this article, alongwith similar equations for stripping
reactions reported earlier [7], complete the set for direct transfer
reactions involving general multishell target states. It is hoped that the
sum rule analysis can now be extended to much larger experimental data to
extract useful information about effective two-body interaction.
\eject \vfill
\newpage
\subsection*{ References}
\small
\begin{enumerate}
\item R. K. Bansal and J. B. French, Phys. Lett. ~{\bf 19}~(1965)~223.
\item J.B.French, in: Many body description of nuclear structure and
reactions, ed. C.Bloch (Academic, New York and London, 1966)~p. 278. 
\item R. K. Bansal and J. B. French, Phys. Lett.~{\bf 11}(1964)~145.
\item R.K.Bansal, Phys. Lett.~{\bf B~40}~(1972)~189.
\item R. K. Bansal and S.Shelly, Phys. Rev. {\bf C~8}~(1973)~282.
\item R. K. Bansal and Ashwani Kumar, Pramana - J. Phys.~{\bf 9}~(1977)~273.
\item R. K. Bansal and Ashwani Kumar, Pramana - J. Phys.~{\bf 32}~(1989)~341.
\item J. B. French and M.H. Macfarlane, Nucl. Phys.~{\bf 26}~(1961)~168.
\item A. H. Wapstra and N. B. Gove, Nucl. Data Tables {\bf 9}~(1961)~265.
\item O. Hansen et al., Nucl. Phys.~{\bf A243}~(1975)~100.
\item O. Hansen et al., Nucl. Phys.~{\bf A253}~(1975)~380.
\item J. L. Yntema, Phys. Rev.~{\bf 186}~(1969)~1144.
\item J. Bommer et al., Nucl. Phys.~{\bf A160}~(1971)~577.
\item S. M Smith et al., Nucl. Phys.~{\bf A113}~(1968)~303.
\item G. Brown et al., Nucl Phys.~{\bf A225}~(1974)~267.
\item P. Martin et al., Nucl. Phys.~{\bf A185}~(1972)~465.
\item J. J. Schwartz and W. Parker Alford,  Phys. Rev.~{\bf 149}~(1966)~820.
\item J. Rapaport et al., Phys. Rev.~{\bf 151}~(1966)~939.
\item G. Mairle et al., Nucl. Phys.~{\bf A134}~(1969)~180.
\item J. H. Bjerregaard, Ole Hansen and G. Sidenius, Phys.
Rev.~{\bf138}~(1965)~B1097. 
\item B. Cujec and I. M. Szoghy, Phys. Rev.~{\bf 179}~(1969)~1060.
\item P. J. Plauger and E. Kashy, Nucl. Phys.~{\bf A152}~(1970)~609.
\item M. S. Chowdhury and H. M. Sen Gupta, Nucl. Phys.~{\bf
A229}~(1974)~484. 
\item J. R. Erskine et al., Phys. Rev.~{\bf 142}~(1966)~633.
\item D. Bachner et al., Nucl. Phys.~{\bf A106}~(1968)~577.
\item M. N. Rao et al., Nucl. Phys.~{\bf A151}~(1970)~351.
\item G. Brown et al., Nucl. Phys.~{\bf A183}~(1972)~471.
\item H. Ohnuma, Phys. Rev.~{\bf C3}~(1971)~1192.
\item A. E. MacGregor and G. Brown, Nucl. Phys.~{\bf A190}~(1972)~548.
\item P. David et al., Nucl. Phys.~{\bf A128}~(1969)~47.
\item M. A. Basher et al., Phys. Rev.~{\bf C 45}~(1992)~1575.
\item J. C. Legg and E. Rost, Phys. Rev.~{\bf 134}~(1964)~B752.
\item T.T.S Kuo and G.E.Brown, Nucl.Phys. {\bf A114} (1968) 241.
\item K. Lips and M. T. McEllistren, Phys. Rev. {\bf C1} (1970) 1009.
\item P. Federman and S. Pittel, Nucl. Phys. {\bf A155} (1970) 161.
\item J. P. Schiffer and W. W. True, Rev. Mod. Phys. {\bf 48} (1976) 191.
\end{enumerate}
\eject \vfill
\newpage
\normalsize
\begin{center}{ \bf Table 1.} List of experiments from which the centroids
have been obtained for the present study and the various combinations of
stripping and pick-up reactions used in calculations. 
\end{center}
\begin{tabular}{l  l  l  c  r  r  r}  \hline
Target & Stripping & Centroid & & Pick-up & Centroid &\\
\cline{3-4}\cline{6-7} 
& Reaction & Isospin& Value(MeV) & Reaction & Isospin & Value(MeV) \\ 
& [Reference] & & & [Reference] & & \\ \hline 
$^{41}$Ca& $(d,p)$~[10] & $T_>$ & 2.819 & $(d,t)$~[11] & $T_<$ & 0.083\\ 
$^{42}$Ca& $(d,p)$~[15] & $T_>$& 0.066 & $(d,t)$~[12] & $T_<$ & 0.000\\ 
$^{42}$Ca& $(^{3}$He,$d)$~[13] & $T_<$&1.079 & $(p,d)$~[16] & $T_<$ &
0.567\\ 
$^{42}$Ca& $(^{3}$He,$d)$~[13] & $T_<$& 1.079 & $(d,t)$~[12] & $T_<$&
0.000\\ 
$^{42}$Ca& $(^{3}$He,$d)$~[13] & $T_>$& 4.234 & $(d,t)$~[12] & $T_<$&
0.000\\ 
$^{44}$Ca& $(d,p)$~[15] & $T_>$& 0.352 & $(p,d)$~[14] & $T_<$& 0.000\\ 
$^{44}$Ca& $(^{3}$He,$d)$~[17] & $T_<$& 0.453 & $(p,d)$~[16] & $T_<$&
0.000\\ 
$^{45}$Sc& $(d,p)$~[18] & $T_>$& 0.383 & $(d,^{3}$He$)$~[19]& $T_>$&
1.086\\ 
$^{46}$Ca& $(d,p)$~[20] & $T_>$& 0.000 & $(d,t)$~[12] & $T_<$& 0.000\\ 
$^{46}$Ti& $(^{3}$He,$d)$~[21] & $T_<$& 0.150 & $(p,d)$~[22] & $T_<$&
0.434\\ 
$^{46}$Ti& $(d,p)$~[23] & $T_>$& 0.555 & $(p,d)$~[22] & $T_>$& 4.760\\ 
$^{48}$Ca& $(^{3}$He,$d)$~[24] & $T_<$& 0.000 & $(p,d)$~[16] & $T_<$&
0.050\\ 
$^{48}$Ti& $(^{3}$He,$d)$~[25] & $T_<$& 0.151
&$(^{3}$He,$\alpha)$~[26]&$T_<$& 0.819\\ 
$^{48}$Ti& $(d,p)$~[27] & $T_>$& 0.583 & $(d,^{3}$He$)$~[28]& $T_>$&
0.000\\ 
$^{50}$Cr& $(d,p)$~[29] & $T_>$& 0.302 & $(t,\alpha)$~[25]&$T_>$& 0.490\\ 
$^{50}$Cr& $(^{3}$He,$d)$~[21]& $T_<$& 0.312
&$(^{3}$He,$\alpha)$~[30]&$T_<$& 0.257\\ 
$^{51}$V& $(^{3}$He,$d)$~[31] & $T_<$& 2.279 & $(p,d)$~[32] & $T_<$&
1.769\\ \hline 
\end{tabular}
\vskip 1.5truecm
{\bf Table 2.} Average two-body interaction parameters for $1f_{7\over2}$
shell nuclei. 
\vskip .5truecm
\begin{tabular}{c c c c} \hline
             &    $\overline{W}^{T=1}_{f_{7\over2}f_{7\over2}}(MeV)$
             &     $\overline{W}^{T=0}_{f_{7\over2}f_{7\over2}}(MeV)$
             &rms deviation\\ \hline
Present calc.     &     $-$0.212         &          $-$1.662   & 0.542\\
Previous calc.[7]  &   $-$0.215        &           $-$1.714  &0.191\\
Kuo-Brown[33]     &   $-$0.128            &            $-$1.154 & -\\
Lips \& McEllistren[34]   &  $-$0.240   &             -& -\\
Federman \&  Pittel[35]    &  $-$0.228   &              -  & -\\
Schiffer \& True [36]    &  -      &
${-1.739}^a$,${-1.594}^{b}$ & -\\  \hline
\end{tabular}\\
{\small $^a$ Derived from a potential fitted to the experimental data}\\
{\small $^b$ Determined from a direct fit to experimental data }
\end{document}